\documentstyle[preprint,aps]{revtex}
\tighten
\newcommand{\bc}{\begin{center}}
\newcommand{\ec}{\end{center}}
\newcommand{\beqn}{\begin{equation}}
\newcommand{\eeqn}{\end{equation}}
\newcommand{\barr}{\begin{eqnarray}}
\newcommand{\earr}{\end{eqnarray}}
\def\neqn {\nonumber}

\def\del{\partial}

\def\eg {{\it e.g}. }
\def\ie {{\it i.e}. }

\def\tr {\mbox{tr}}
\def\ctr {\mbox{Tr}}

\def\PL #1 #2 #3 {Phys. Lett.~{\bf#1} (#2) #3}
\def\NP #1 #2 #3 {Nucl. Phys.~{\bf#1} (#2) #3}
\def\NPP #1 #2 #3 {Nucl. Phys.~{\bf B} (Proc.~Suppl.)~{\bf#1} (#2) #3}
\def\ZP #1 #2 #3 {Z.~Phys.~{\bf#1} (#2) #3}
\def\PR #1 #2 #3 {Phys. Rev.~{\bf#1} (#2) #3}
\def\PP #1 #2 #3 {Phys. Rep.~{\bf#1} (#2) #3}
\def\PRL #1 #2 #3 {Phys. Rev.~Lett.~{\bf#1} (#2) #3}
\def\PTP #1 #2 #3 {Prog. Theor.~Phys.~{\bf#1} (#2) #3}
\def\MPL #1 #2 #3 {Mod. Phys.~Lett.~{\bf#1} (#2) #3}
\def\IJM #1 #2 #3 {Int. J.~Mod.~Phys.~{\bf#1} (#2) #3}

\newlength{\minitwocolumn}
\renewcommand{\thefigure}{\arabic{figure}} 
\newcommand{\fcaption}[1]{\refstepcounter{figure} Fig.~\thefigure  #1}
\input epsf
\begin{document}
\begin{flushright}
YITP-97-35 \\
June 1997
\end{flushright}
\vskip 2cm
\bc
\Large\bf
${\rm U_{A}}(1)$ Anomaly in Background Fields \\
Dominated by QCD-Monopoles on SU(2) Lattice
\vskip 1cm
\ec
\centerline{\large
Shoichi~SASAKI$^{\;a)}$\footnote[3]{E-mail address:~ssasaki@yukawa.kyoto-u.ac.jp}
 and Osamu~MIYAMURA$^{\;b)}$}
\vskip 0.5cm

\centerline{\em a) Yukawa Institute for Theoretical Physics, Kyoto University}
\vskip 0.2cm
\centerline{\em b) Department of Physics, Hiroshima University}
\baselineskip 24pt
\vskip 1cm
\begin{abstract}
\baselineskip 20pt
\indent
We study ${\rm U}_{\rm A}(1)$ anomaly of non-perturbative QCD
in the maximally abelian gauge on SU(2) lattice.
The existence of the strong correlation between
QCD-monopoles and instantons in the abelian gauge
is shown by both analytic
and numerical works including lattice simulations.
These results bring us a conjecture that the ${\rm U}_{\rm A}(1)$
symmetry would be explicitly broken in the background
fields dominated by QCD-monopoles.
We find an evidence for our conjecture by measuring
the chiral-asymmetric zero modes of the Dirac operator
in the backgrounds of QCD-monopoles.

\end{abstract}
\newpage

\indent
As for the appearance of color-magnetic monopoles in ${\rm SU}(N_{c})$
gauge theory, 't Hooft proposed an interesting idea of the 
abelian gauge fixing \cite{Hooft2}, 
which is defined by the gauge transformation in
the coset space of the gauge group in order to fix the gauge degrees
of freedom up to the maximally abelian subgroup.
In the abelian gauge, point-like singularities in the 
three-dimensional space ${\bf R}^3$ under the maximally abelian subgroup
can be identified as color-magnetic monopoles \cite{Hooft2}, 
which will be called QCD-monopoles hereafter.
In other words, QCD-monopoles originate from the same topological nature 
as the 't Hooft-Polyakov monopoles, which correspond
to the homotopy group $\pi_{2}({\rm SU}(N_{c})/{\rm U}(1)^{N_{c}-1})
=Z^{N_{c}-1}_{\infty}$ \cite{Raja}.
However, QCD-monopoles have no condition to exist as classically stable 
and/or finite-energy
solutions, unlike the 't Hooft-Polyakov monopoles are the explicit solutions 
of the field equation \cite{Raja}. 
One thus conjectures that the realistic model of QCD vacuum would 
be characterized by the highly quantum feature of QCD-monopoles, \eg
its condensation, rather than the classical one \cite{Hooft1}.
Then the dual Meissner effect, which yields the exclusion of the 
color-electric fields, must be realized \cite{Hooft1}.
The recent lattice QCD simulations \cite{Kronfeld}-\cite{Kitahara}
support this conjecture that 
QCD-monopoles play a crucial role on color confinement through their 
condensation\footnote[2]{QCD-monopole condensation is characterized by 
the presence of the long and tangled monopole trajectories in the 
four-dimensional space ${\bf R}^4$ and can be interpreted as 
the Kosterlitz-Thouless type 
phase transition \cite{Kitahara}.}.

As well known, QCD has also classical and non-trivial 
gauge configurations, \ie instantons as topological defects
in the Euclidean space ${\bf R}^4$ 
corresponding to the homotopy group 
$\pi_{3}({\rm SU}(N_{c}))=Z_{\infty}$ \cite{Raja}.
It seems that instantons and QCD-monopoles are thought to be 
hardly related to each other since these topological objects appear 
from different non-trivial homotopy groups.
However, the recent analytical works 
have demonstrated 
the QCD-monopole as a classically stable solution 
in the background fields of the instanton configuration using
the abelian gauge fixing 
\cite{{Suganuma2},{Suganuma3},{Suganuma4},{Chernodub},{Brower}}. 
Furthermore, 
the several lattice QCD simulations have
shown the existence of the strong correlation between
instantons and QCD-monopoles in the highly quantum vacuum 
\cite{{Miyamura1},{Suganuma4},{Markum},{Suganuma5}}
as well as the semi-classical vacuum \cite{{Brower},{Teper},{Sasaki}}.

Here, we remind that instantons are important topological objects in QCD 
relating to the ${\rm U_{A}}(1)$ problem \cite{Hooft3}. It is well known that
the index of the massless Dirac operator
in the instanton background fields is equal to the Pontryagin index,
\ie the topological charge $Q$:
%
%
\beqn
n_{+} - n_{-} = Q\;\;\;,
\eeqn
where $n_{+}$ ($n_{-}$) is the number of zero modes with the positive 
(negative) chirality. 
The previous relation is well known as the Atiyah-Singer index theorem.
The index of the Dirac operator corresponds to the number of 
chiral-asymmetric zero modes that yield the chiral anomaly in 
the global ${\rm U}_{\rm A}(1)$ symmetry.
By the existence of this anomaly,
the ${\rm U}_{\rm A}(1)$ symmetry is regarded as explicitly 
broken at the quantum level.
This mechanism plays an essential role on the
resolution of the ${\rm U}_{\rm A}(1)$ problem \cite{Hooft3}.

In this paper, 
we aim to reexamine the relation
between QCD-monopoles and instantons 
through ${\rm U}_{\rm A}(1)$ anomaly.
By using the Monte Carlo simulation
on SU(2) lattice, we measure the topological charge and the zero 
eigenvalues of the Dirac operator in both the ``monopole dominating'' and
``monopole absent'' background fields.

The Maximally Abelian (MA) gauge fixing \cite{Hooft2} 
was advocated by 't Hooft in order to define the magnetic monopole 
in the renormalizable and the Lorentz 
invariant way in the continuum,
$(\del_{\mu}\pm igA_{\mu}^{3})A_{\mu}^{\pm}=0$
where $A_{\mu}^{\pm}=A_{\mu}^{1}\pm iA_{\mu}^{2}$.
In the lattice formulation \cite{Kronfeld}, this gauge fixing is expressed by 
diagonalizing the 
following operator $X(n)$ through the gauge transformation 
$U_{\mu}(n) \rightarrow V(n)U_{\mu}(n) V^{\dag}(n+{\hat \mu})$,
%
%
\beqn
X(n) = \sum_{\mu}\left\{ U_{\mu}(n) \sigma_{3} U^{\dag}_{\mu}(n)
+ U^{\dag}_{\mu}(n-{\hat \mu}) \sigma_{3} U_{\mu}(n-{\hat \mu})
\right\} \;\;,
\eeqn
where $U_{\mu}(n)$ are link variables.
While $X(n)=X^{1}(n)\sigma_{1}+X^{2}(n)\sigma_{2}+X^{3}(n)\sigma_{3}$,
this gauge fixing means that the off-diagonal elements are locally minimized 
in every sites by the gauge transformation, \ie $X^{1}(n)=X^{2}(n)=0$.
Instead of this procedure, the gauge transformation is actually carried out by 
maximizing the gauge dependent variable $R$ \cite{Kronfeld},
%
%
\beqn
R=\sum_{n,\;\mu}\tr\left\{
\sigma_{3} U_{\mu}(n) \sigma_{3} U^{\dag}_{\mu}(n) \right\}\;\;.
\eeqn
Maximizing $R$ is equivalent to making $X(s)$ diagonal at all sites.

Once the gauge transformation is done by the above procedure, we  
factorize the SU(2) link variable $U_{\mu}(n)$ into the 
abelian link variable $u_{\mu}(n)$ and off-diagonal part $M_{\mu}(n)$ 
\cite{Kronfeld} as
%
%
\beqn
U_{\mu}(n)=M_{\mu}(n)\cdot u_{\mu}(n) \;\;,
\eeqn
where
%
%
\barr
u_{\mu}(n)&\equiv&\exp \{ i\sigma_{3}\theta_{\mu}(n) \} \;\;, \\
M_{\mu}(n)&\equiv&\exp \{ i\sigma_{1}
C^{1}_{\mu}(n)+i\sigma_{2}C^{2}_{\mu}(n) \} \;\;.
\earr
Here, $\theta_{\mu}(n)$ is the ${\rm U}(1)$ gauge field and
$C^{1}_{\mu}(n)$ and $C^{2}_{\mu}(n)$ correspond to 
charged matter fields under a residual ${\rm U}(1)$ gauge transformation.
Performing the ${\rm U}(1)$ gauge transformation 
on the original SU(2) link variable, $u_{\mu}(n)$ and $M_{\mu}(n)$ 
are transformed \cite{Kronfeld} as  
%
%
\barr
u_{\mu}(n)&\longrightarrow& u_{\mu}'(n)=d(n)u_{\mu}(n)
d^{\dag}(n+{\hat \mu}) \;\;,\\
M_{\mu}(n)&\longrightarrow& M_{\mu}'(n)=d(n)M_{\mu}(n)d^{\dag}(n) 
\;\;,
\earr
where $d(n)=\exp\{i\sigma_{3}\varphi(n)\}$.
In this way, $u_{\mu}(n)$ and $M_{\mu}(n)$ behave like 
an abelian `photon' and an adjoint `matter' field, respectively.

Our next task is to look for the magnetic monopole in terms of
the ${\rm U}(1)$ variables. We consider the product of 
${\rm U}(1)$ link variables
around an elementary plaquette,
%
%
\beqn
u_{\mu \nu}(n)=u_{\mu}(n) u_{\nu}(n+{\hat \mu})
u^{\dag}_{\mu}(n+{\hat \nu}) u^{\dag}_{\nu}(n)=e^{i\sigma_{3}
\theta_{\mu \nu}} \;\;,
\eeqn
where the abelian field strength $\theta_{\mu \nu}(n)\equiv
\theta_{\nu}(n+{\hat \mu})-\theta_{\nu}(n)-\theta_{\mu}(n+{\hat \nu})
+\theta_{\mu}(n)$.
It should be noted that the ${\rm U}(1)$ plaquette variable 
is a multiple valued 
function as the abelian field strength due to the compactness of
the residual ${\rm U}(1)$ gauge group. 
Then we can divide the abelian field strength into two parts as
%
%
\beqn
\theta_{\mu \nu}={\bar \theta}_{\mu \nu}+2\pi
N_{\mu \nu} \;\;,
\eeqn
where ${\bar \theta}_{\mu \nu}$ is the regular part defined in 
$-\pi <{\bar \theta}_{\mu \nu}\leq \pi$ and 
$N_{\mu \nu}\in {\bf Z}$ is the modulo $2\pi$ of $\theta_{\mu \nu}$.
Here, it is known that the SU(2) link variable behaves as $U_{\mu}\simeq 
u_{\mu}$ in the MA gauge \cite{Hioki}. 
In this sense, ${}^{\ast}N_{\mu \nu}\equiv 
{1 \over 2}\varepsilon_{\mu \nu \rho \sigma}N_{\rho \sigma}$ 
corresponds to the Dirac string following the DeGrand-Toussaint's definition
in the compact QED \cite{DeGrand}. Then, monopole currents 
$k_{\mu}(n)$ are identified as topological conserved currents defined by 
$k_{\mu}(n)=\del_{\nu}{}^{\ast}N_{\mu \nu}(n+{\hat \mu})$ \cite{DeGrand}. 

The next aim is to extract the 
contribution of the monopole dominated part from the abelian link variable.
First, we define the two abelian gauge fields using two parts of
the abelian field strength \cite{Miyamura2} as below,
%
%
\barr
\theta^{\rm Ph}_{\mu}(n) &\equiv& \sum_{m}G(n-m)\del_{\lambda}
{\bar \theta}_{\lambda \mu}(m) \;\;,\\
\theta^{\rm Ds}_{\mu}(n) &\equiv& 2\pi\sum_{m}G(n-m)\del_{\lambda}
N_{\lambda \mu}(m) \;\;,
\earr
where $G(n-m)$ is the lattice Coulomb propagator.
$\theta^{\rm Ph}_{\mu}$, which is called `regular part', is composed 
of the regular part of the abelian field strength ${\bar \theta_{\mu 
\nu}}$ \cite{Miyamura2}.
On the other hand,  $\theta^{\rm Ds}_{\mu}$, which is called `singular 
part', is composed of the Dirac string part of the 
the abelian field strength \cite{Miyamura2}.
It is noted that the sum of two values is the original 
abelian gauge field in the Landau gauge, $\del_{\mu}
\theta^{L}_{\mu}(n)=0$ \cite{Miyamura1},
%
%
\beqn
\theta^{\rm Ph}_{\mu}(n)+\theta^{\rm Ds}_{\mu}(n)
=\sum_{m}G(n-m)\del_{\lambda}
\theta_{\lambda \mu}(m)=\theta^{L}_{\mu}(n) \;\;,
\eeqn
where a superscript $L$ denotes the Landau gauge.
This procedure could correspond to dividing 
a `regular photon' part from a `monopole' part \cite{Miyamura2}.

The corresponding SU(2) variables are reconstructed from $\theta^{\rm 
Ph}_{\mu}$ and $\theta^{\rm Ds}_{\mu}$ by multiplying the off-diagonal
factor $M_{\mu}$ \cite{Miyamura1}-\cite{Suganuma4} as
%
%
\barr
U^{\rm Ph}_{\mu}(n) &\equiv& M_{\mu}(n)\exp\{i\sigma_{3}
\theta^{\rm Ph}_{\mu}(n)\} \;\;,\\
U^{\rm Ds}_{\mu}(n) &\equiv& M_{\mu}(n)\exp\{i\sigma_{3}
\theta^{\rm Ds}_{\mu}(n)\} \;\;.
\earr
Here, the fact that $U_{\mu}=U^{\rm Ph}_{\mu}\cdot u^{\rm Ds}_{\mu}
=U^{\rm Ds}_{\mu}\cdot u^{\rm Ph}_{\mu}$ in the Landau gauge 
should be kept in mind. Thus, we shall use $U^{\rm Ds}_{\mu}$ as the 
`monopole dominating' SU(2) link variable and $U^{\rm Ph}_{\mu}$ as
the `monopole absent' SU(2) link variable \cite{Miyamura1}-\cite{Suganuma4}.

Next, we see how instantons are defined 
in the lattice formulation of QCD.
Of course, in discretised space-time, we lose inherently the topology 
in the strict mathematical sense. Nevertheless, we expect that the 
topological character could be neatly identified near the continuum 
limit, since the variation of fields becomes smoother than 
the size of the lattice spacing \cite{Luscher}.
We use the simplest expression for the topological charge \cite{Rossi} as
%
%
\beqn
Q_{L}={1 \over 32\pi^{2}}\sum_{n}
\varepsilon_{\mu \nu \rho \sigma}
{\rm Tr} \{ U_{\mu \nu}(n)U_{\rho \sigma}(n) \} \;\;,
\eeqn
where $U_{\mu \nu}(n)$ is the plaquette variable.
In the naive continuum limit,  
${\rm Tr} \{ U_{\mu \nu}(n)U_{\rho \sigma}(n) \}$ is reduced to 
${\rm Tr} \{a^{4}g^{2}G_{\mu \nu}G_{\rho \sigma} + O(a^{5})\}$ 
\cite{Rossi}.
The value $Q_{L}$ has not only $O(a^{2})$ corrections, but also renormalized 
multiplicative corrections of $O(a^{0})$ \cite{Giacomo}. 
Consequently, $Q_{L}$ is not an integer except for the continuum limit.
However, it is known that the topological feature of $Q_{L}$ can be 
extracted by removing the short wavelength fluctuations of 
the field configuration by using the following procedure. 
We use the cooling method \cite{Ilgenfritz}, 
in which each link variable $U_{\mu}$ is replaced by
%
%
\beqn
{\tilde U}_{\mu}(n)=c \sum_{\mu \perp \nu}U_{\nu}(n)U_{\mu}(n+{\hat \nu})
U^{\dag}_{\nu}(n+{\hat \mu}) \;\;.
\eeqn
where a factor $c$ ensures that a new link variable ${\tilde U}_{\mu}$ is the 
element of SU(2) group. After this procedure, 
which is called a {\it cooling sweep}, the action is reduced; 
$S[{\tilde U}_{\mu}] < S[U_{\mu}]$. 
As a consequence, the field configuration locally smoothened. 
In Fig.\ref{fig:Cool_Ds}-\ref{fig:Cool_Ph}, the cooling curves for the 
topological charge; $Q_L$, the integral of the absolute value of the 
topological density; $I_Q$\footnote[2]{This quantity is defined as 
$I_Q = {1 \over 32\pi^2}\sum_{n} \varepsilon_{\mu \nu \rho \sigma}
|\ctr\{U_{\mu \nu}(n)U_{\rho \sigma}(n)\}|$ corresponding to
the total number of topological pseudoparticles (instantons and
anti-instantons).} 
and the action divided by $8\pi$; $\tilde S$ are shown
as typical examples.

In order to examine the eigenvalue of the Dirac operator on the 
lattice, we adopt the Wilson fermion operator \cite{Wilson1};
%
%
\barr
\lefteqn{{D}(n,\;m;\;U)} \neqn \\
&&=
\delta_{n,\;m}-\kappa\sum_{\mu}\left[
(r_{\rm w}-\gamma_{\mu})U_{\mu}(n)\delta_{n+{\hat \mu},\;m}
+(r_{\rm w}+\gamma_{\mu})U^{\dag}_{\mu}(n-{\hat \mu})
\delta_{n-{\hat \mu},\;m}\right] \;\;.
\earr
where $r_{\rm w}$ is the Wilson parameter.
We see that the choice $r_{\rm w}=1$ is quite special since
$1\pm \gamma_{\mu}$ are orthogonal projection operators. Thus, we use
$r_{\rm w}=1$ hereafter. In the naive argument, the 
Wilson fermion does not have
the chiral symmetry due to the Wilson term.
However, the effect of chiral symmetry breaking was systematically examined 
through the chiral Ward identities \cite{Testa} 
and the Wilson term is necessary 
to maintain the axial vector anomaly \cite{Smit1}. 
It is known that in the strong coupling region
the ordinary mass term and the Wilson term cancel out
in the pseudo-scalar mass, which is symbolically called the pion mass,
at some $\kappa=\kappa_{c}(\beta)$.
In the strong coupling limit,
the pion mass is zero at
$\kappa_{c}(\beta\rightarrow 0)\simeq{1 \over 4}$ \cite{Kawamoto2}.
In the weak coupling regime, perturbative calculations indicate 
that the mass of the fermion becomes equal to zero 
along the line $\kappa=\kappa_{c}(\beta)$, which 
ends at $\kappa_{c}(\beta \rightarrow 
\infty)\simeq{1 \over 8}$. 
In fact, many Monte Carlo simulations of lattice QCD with
the Wilson fermion have shown the existence of such a critical 
line $\kappa_{c}(\beta)$ in the $\kappa - \beta$ plane, where
the pion mass vanishes.
The partial symmetry restoration would be realized 
near the vicinity of the critical line.

The Dirac operator $D$ defined by the Wilson fermion
loses a feature as the hermite operator owing to the discretization
of the space-time. However, we can easily see that 
the operator $\gamma_{5}{D}$ or ${D}^{\dag}{D}$ is a hermite 
matrix. 
Then, we consider the eigenvalue problems for each
operator by using the Lanczos algorithm.
To discriminate quantum fluctuations
among eigenvalues easily, we use
the link variables smoothening the short wavelength with the
cooling method in the same way as calculations of the topological 
charge \cite{Teper2}.
The presence of chiral-asymmetric zero modes can be found by its sign change
through the variation of the hopping parameter $\kappa$ \cite{Itoh}.
It is actually true that such a eigenvalue spectrum of $\gamma_{5}{D}$
inherently coincides with the operator $D$'s one at $\kappa = \kappa_{c}$, 
where the eigenvalue spectrum crosses a zero line. 

We measure two sets of quantities, \ie the eigenvalue 
spectrum of $\gamma_{5}{D}$
and the topological charge $Q_{L}$, by using `monopole dominating' 
SU(2) link variable $U^{\rm Ds}_{\mu}$ and `photon dominating 
(monopole absent)' 
SU(2) link variable $U^{\rm Ph}_{\mu}$ in the MA gauge on an $8^{4}$ lattice 
with $\beta=2.4$.
Fig.\ref{fig:Zero_Ds1}-\ref{fig:Zero_Ds8} 
show the low-lying spectra of 
$\gamma_{5}{D}$ at various values of $\kappa$ for 8 configurations
with 20 cooling sweeps in the quenched approximation.
It is clear that there exist chiral-asymmetric zero modes around $\kappa\simeq 
0.132$ in the background fields of the `monopole dominating' part.
Furthermore, all 8 configurations in the background fields dominated 
by monopoles hold an analogue of ``the Atiyah-Singer 
index theorem''; $n_{+}-n_{-} \simeq Q_{L}$
where $n_{+}$($n_{-}$) is the number of zero modes with positive 
(negative) ``chirality'' defined by the sign of its eigenvalue in the 
limit $\kappa \uparrow \kappa_{c}$.
On the other hand, Fig.\ref{fig:Zero_Ph1}-\ref{fig:Zero_Ph8} 
show that
no existence of chiral-asymmetric zero modes is found in background
fields of the `photon dominating (monopole absent)' part in all 
8 configurations.
In this case, each topological charge is also 
equal to zero \cite{Suganuma3}-\cite{Miyamura1}.
By using 50 configurations, 
we also examine the eigenvalue of
a positive-definite hermitian operator ${D}^{\dag}{D}$ 
associated with the Dirac operator $D$, in the `monopole dominating' 
background fields and the `photon dominating' background fields. 
There certainly exist almost zero modes 
in the `monopole dominating' background,
though we can not find the corresponding zero modes 
in the `photon dominating' background.
Therefore, in the `monopole dominating' fields,
the explicit breaking of the ${\rm U}_{\rm A}(1)$ symmetry
occurs due to the existence of the chiral-asymmetric zero modes.

In conclusion,
we have investigated the eigenvalue problems for the Dirac
operator, which is defined by the Wilson fermion,
in the background fields of 
the `monopole dominating' (Ds) part and the `photon dominating 
(monopole absent)' (Ph) 
part by using the SU(2) lattice with $8^{4}$ and $\beta = 2.4$.
In only the background fields dominated by QCD-monopoles,
the explicit breaking of the ${\rm U}_{\rm A}(1)$ symmetry
occurs due to the existence of the chiral-asymmetric zero modes. 
We have found the monopole dominance for the ${\rm U}_{\rm 
A}(1)$ anomaly, and also confirmed that ``the Atiyah-Singer index theorem'' 
is satisfied in the backgrounds of QCD-monopoles.

We would like to acknowledge fruitful discussions with H. Suganuma
and H. Toki at Research Center for Nuclear Physics of Osaka
University, where most of the present study has been carried out.
All lattice QCD simulations in this paper have been performed on the
Intel Paragon XP/S(56 node) at the Institute for Numerical Simulations
and Applied Mathematics of Hiroshima University.
One of the authors (S.S.) is supported by Research Fellowships of the
Japan Society for the Promotion of Science for Young Scientists.
%

\newpage
\centerline{\large FIGURE CAPTIONS}

\vspace{0.5cm}

\begin{description}

\item[Fig.1]
\begin{minipage}[t]{13cm}
\baselineskip=20pt
The cooling  curves for $Q_L$, $I_Q$ and $\tilde S$ are examined in 
the 'monopole dominating'(Ds) part and the 'photon dominating 
(monopole absent)'(Ph) part on an $8^4$ lattice with $\beta=2.4$.
We show typical examples in the case of
(a) $Q_L({\rm Ds})\neq 0$, (b) $Q_L({\rm Ds})=0$ and (c) $Q_L({\rm Ph})=0$.
$I_Q({\rm Ds})$ tends to remain finite during the 
cooling process. On the other hand, $I_Q({\rm Ph})$
quickly vanish by only less than 5 cooling sweeps.
Therefore, topological pseudoparticles seem unable to live in the Ph
part, but only survive in the Ds part in the abelian gauge.
\end{minipage}
\vspace{1.0cm}
\item[Fig.2]
\begin{minipage}[t]{13cm}
\baselineskip=20pt
The eigenvalue spectrum of $\gamma_{5}D$ in the `monopole 
dominating' background fields (Ds part) as a function
of the hopping parameter $\kappa$ at 20 cooling sweeps on an $8^{4}$
lattice with $\beta = 2.4$. The chiral-asymmetric zero
modes are found in several configurations, where the topological charges have
nonzero value. 
\end{minipage}
\vspace{1.0cm}
\item[Fig.3]
\begin{minipage}[t]{13cm}
\baselineskip=20pt
The eigenvalue spectrum of $\gamma_{5}D$ in the `photon 
dominating (monopole absent)' background fields (Ph part) as a function
of the hopping parameter $\kappa$ at 20 cooling sweeps on an $8^{4}$
lattice with $\beta = 2.4$.
Both the chiral-asymmetric zero modes and the
nonzero topological charges are not found in each configuration.
\end{minipage}

\end{description}
\newpage
%
{\setcounter{enumi}{\value{figure}}
\addtocounter{enumi}{1}
\setcounter{figure}{0}
\renewcommand{\thefigure}{\arabic{enumi}(\alph{figure})}

%
%
\noindent
\centerline{\epsfysize=6.0cm
\epsfbox{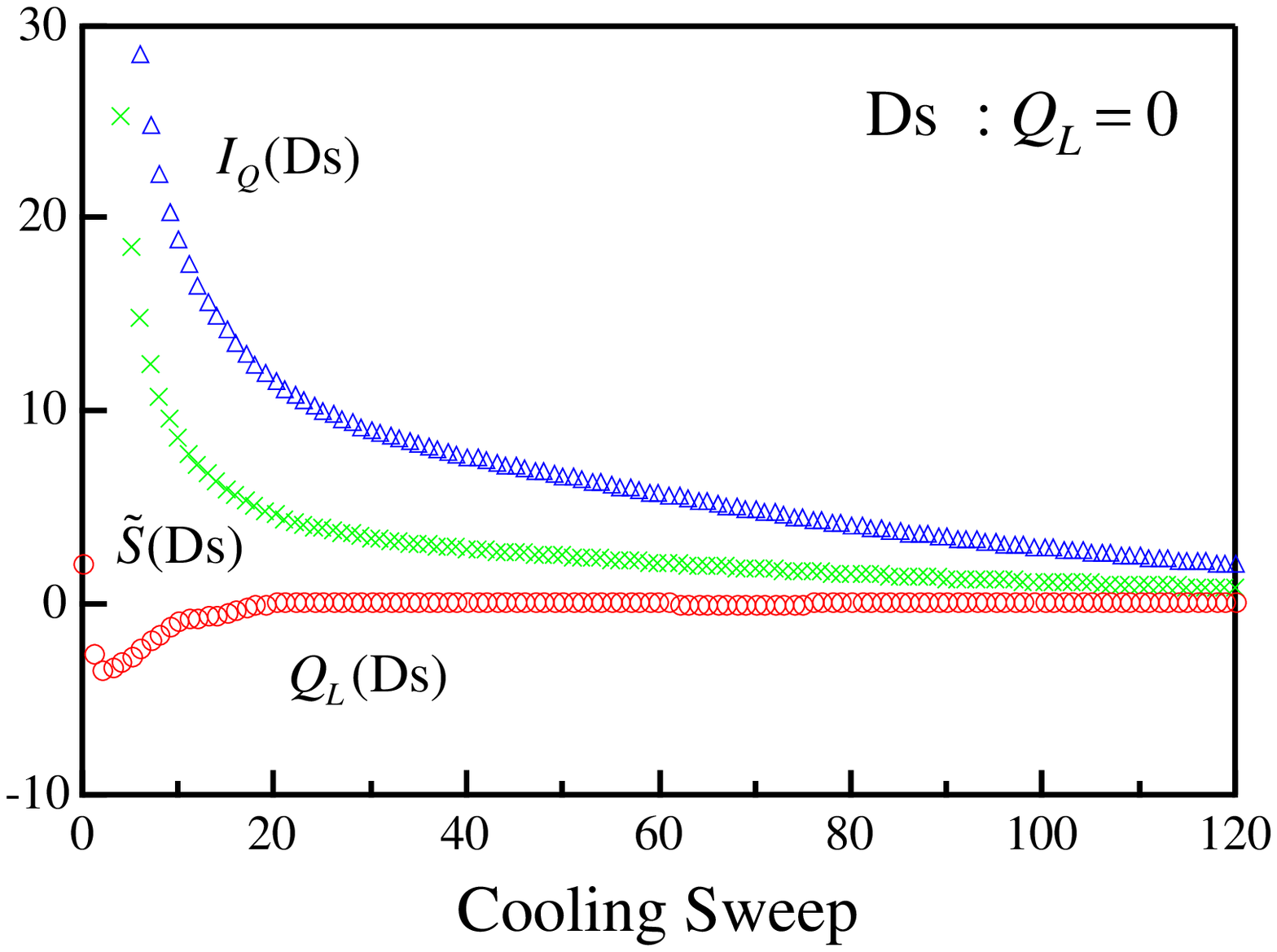}}
\centerline{\fcaption{\label{fig:Cool_Ds}}}

\vspace{0.5cm}
%
%
\noindent
\centerline{\epsfysize=6.0cm
\epsfbox{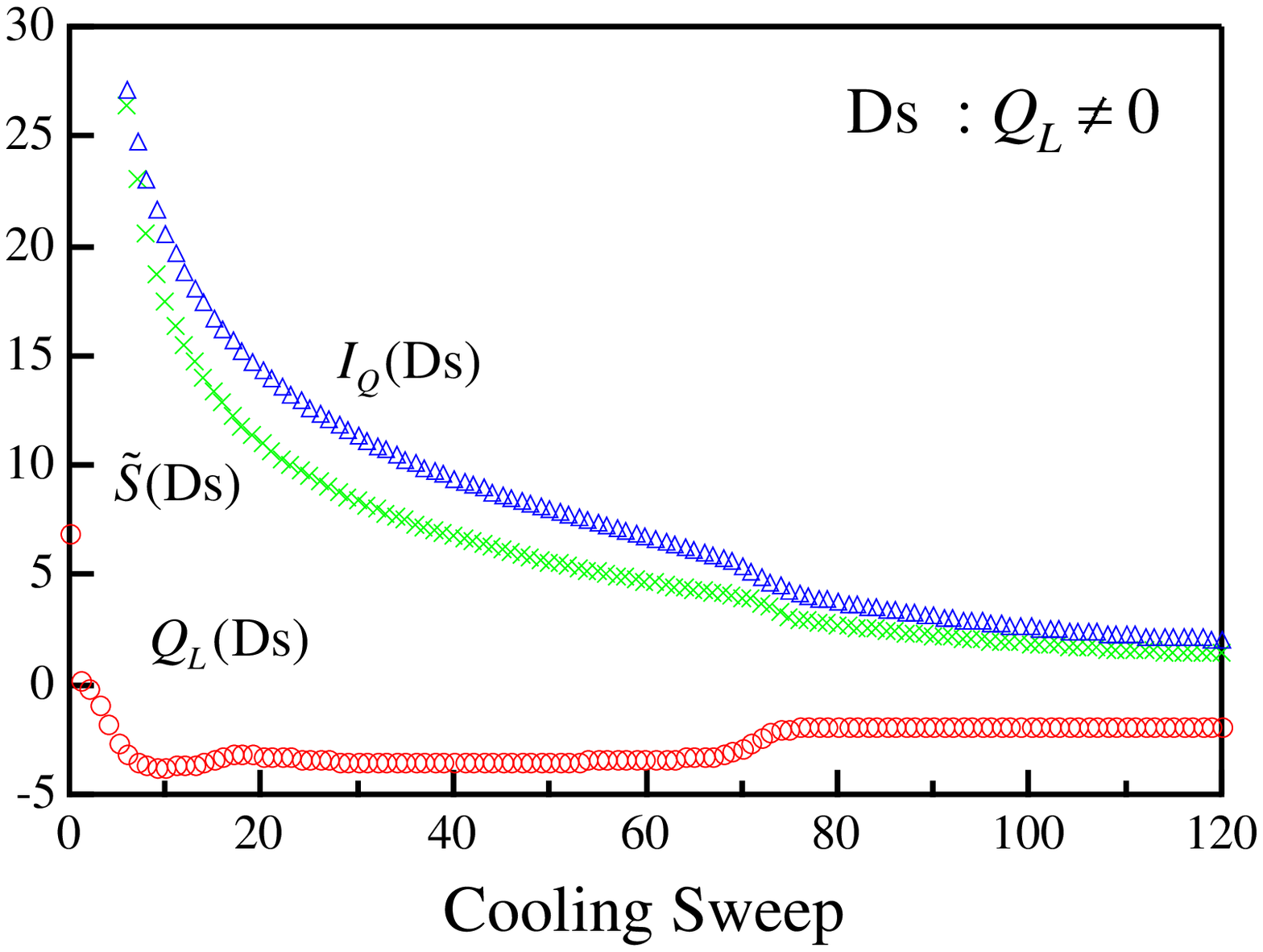}}
\centerline{\fcaption{\label{fig:Cool_Dsz}}}

\vspace{0.5cm}
%
%
\noindent
\centerline{\epsfysize=6.0cm
\epsfbox{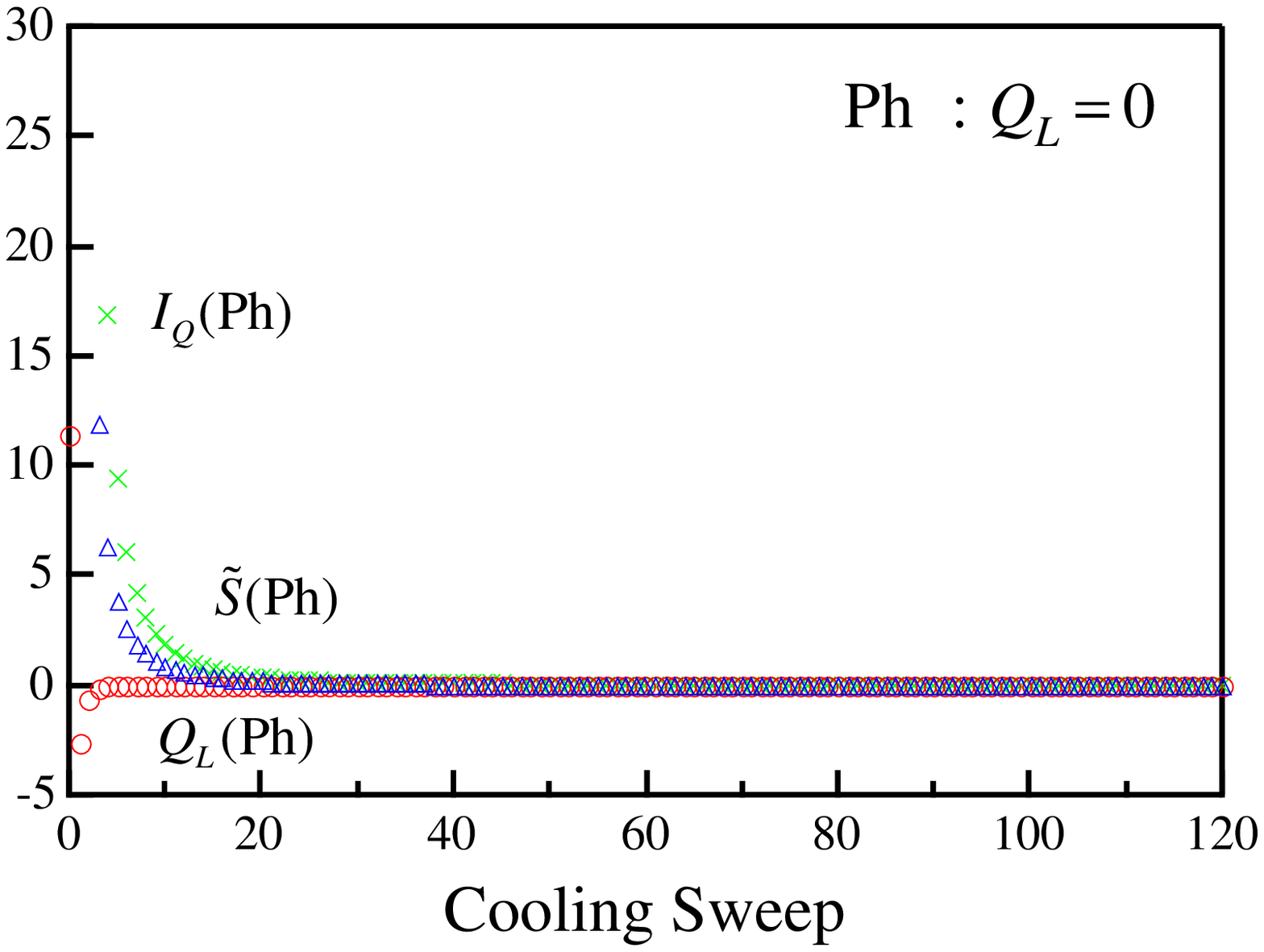}}
\centerline{\fcaption{\label{fig:Cool_Ph}}}
\setcounter{figure}{\value{enumi}}
}
\newpage
\vspace*{1.5cm}
{\setcounter{enumi}{\value{figure}}
\addtocounter{enumi}{1}
\setcounter{figure}{0}
\renewcommand{\thefigure}{\arabic{enumi}(\alph{figure})}

%
%
\noindent
\begin{minipage}{\minitwocolumn}
\centerline{\epsfxsize=7.0cm
\epsfbox{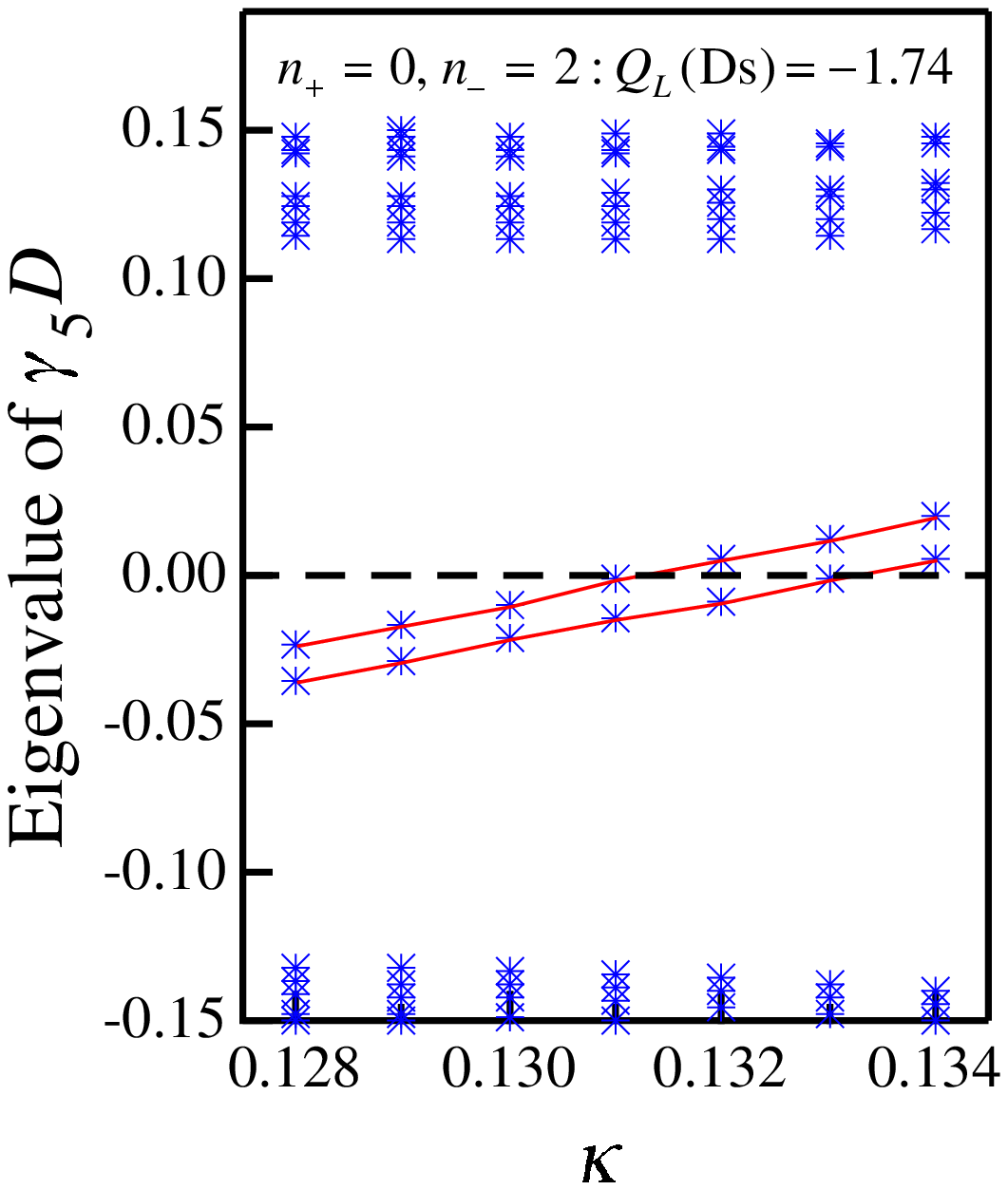}}
\centerline{\fcaption{\label{fig:Zero_Ds1}}}
\end{minipage}
\hspace{\columnsep}
\begin{minipage}{\minitwocolumn}
\centerline{\epsfxsize=7.0cm
\epsfbox{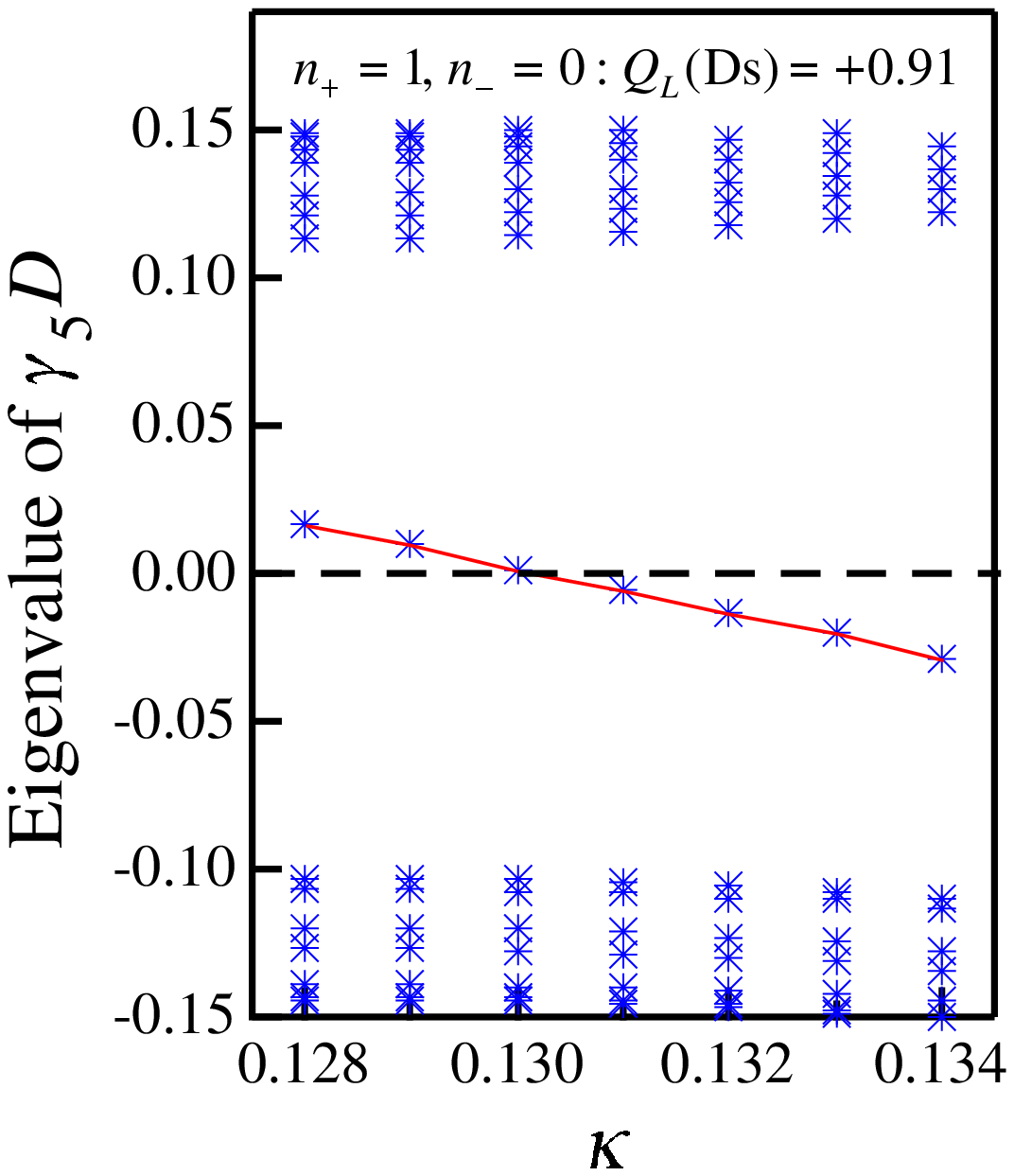}}
\centerline{\fcaption{\label{fig:Zero_Ds2}}}
\end{minipage} 

\vspace{1.5cm}
\noindent
%
%
\begin{minipage}{\minitwocolumn}
\centerline{\epsfxsize=7.0cm
\epsfbox{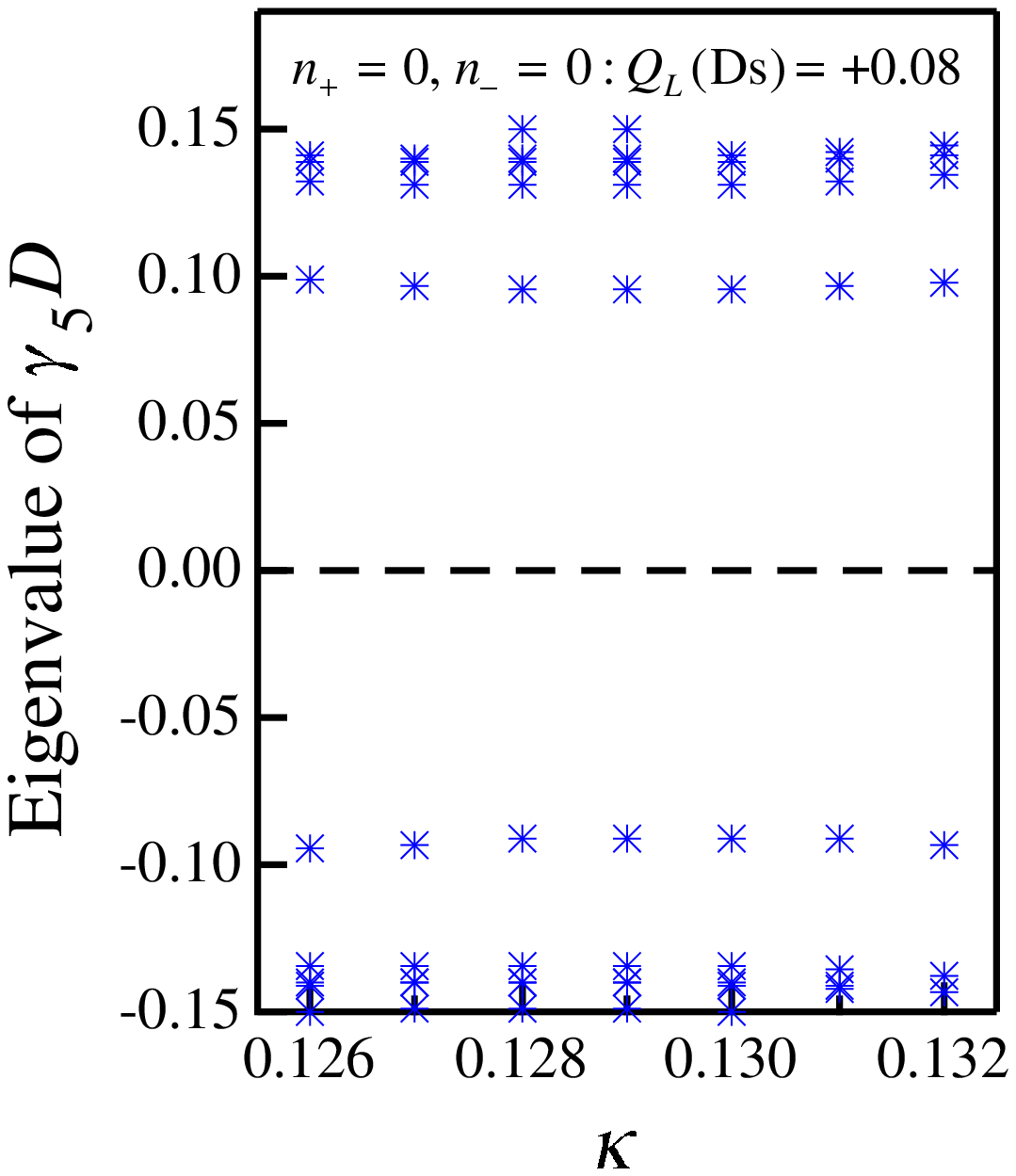}}
\centerline{\fcaption{\label{fig:Zero_Ds3}}}
\end{minipage}
\hspace{\columnsep}
\begin{minipage}{\minitwocolumn}
\centerline{\epsfxsize=7.0cm
\epsfbox{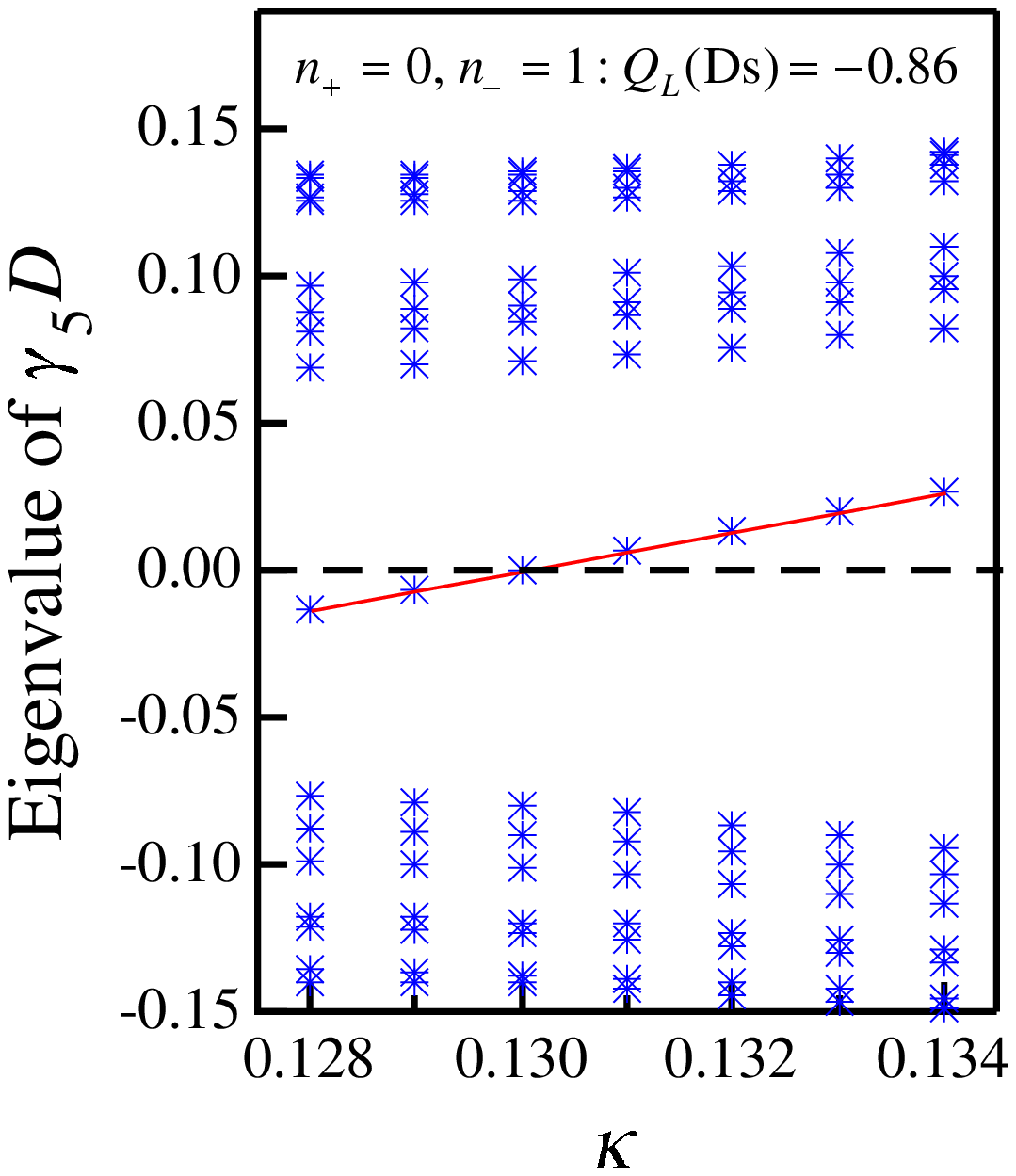}}
\centerline{\fcaption{\label{fig:Zero_Ds4}}}
\end{minipage} 
%
\newpage

\vspace*{1.5cm}
\noindent
%
%
\begin{minipage}{\minitwocolumn}
\centerline{\epsfxsize=7.0cm
\epsfbox{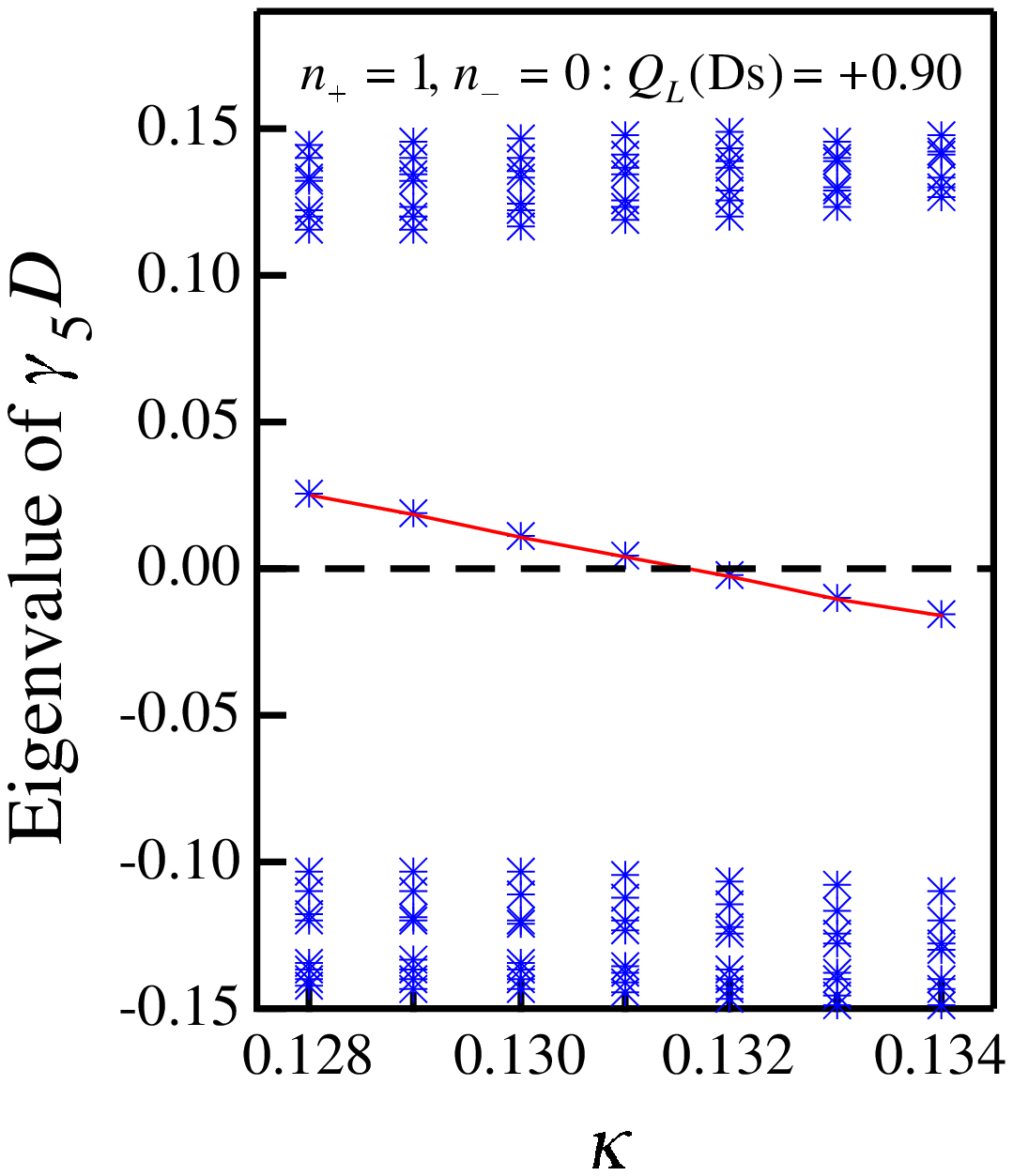}}
\centerline{\fcaption{\label{fig:Zero_Ds5}}}
\end{minipage}
\hspace{\columnsep}
\begin{minipage}{\minitwocolumn}
\centerline{\epsfxsize=7.0cm
\epsfbox{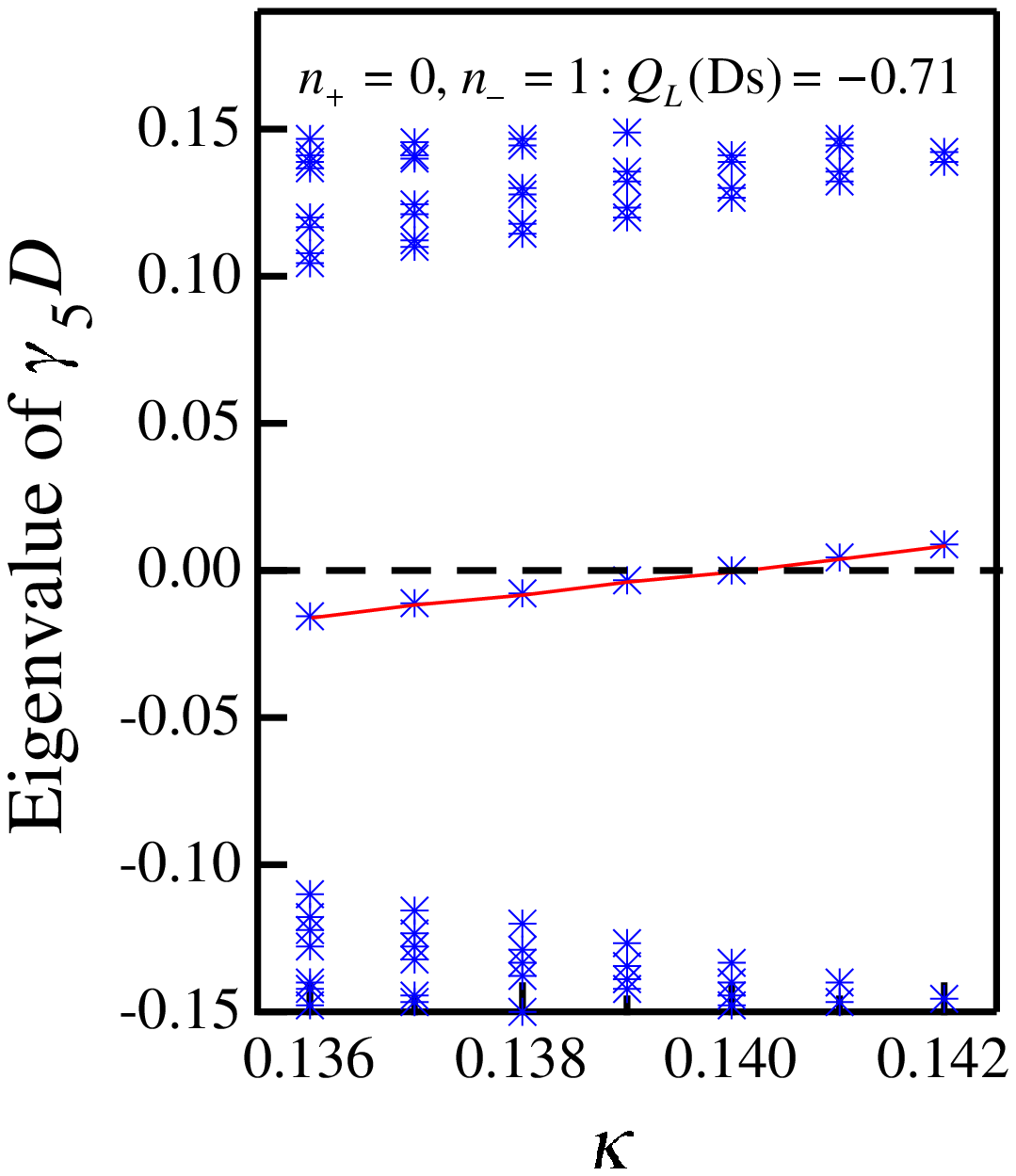}}
\centerline{\fcaption{\label{fig:Zero_Ds6}}}
\end{minipage}

\vspace{1.5cm}
\noindent
%
%
\begin{minipage}{\minitwocolumn}
\centerline{\epsfxsize=7.0cm
\epsfbox{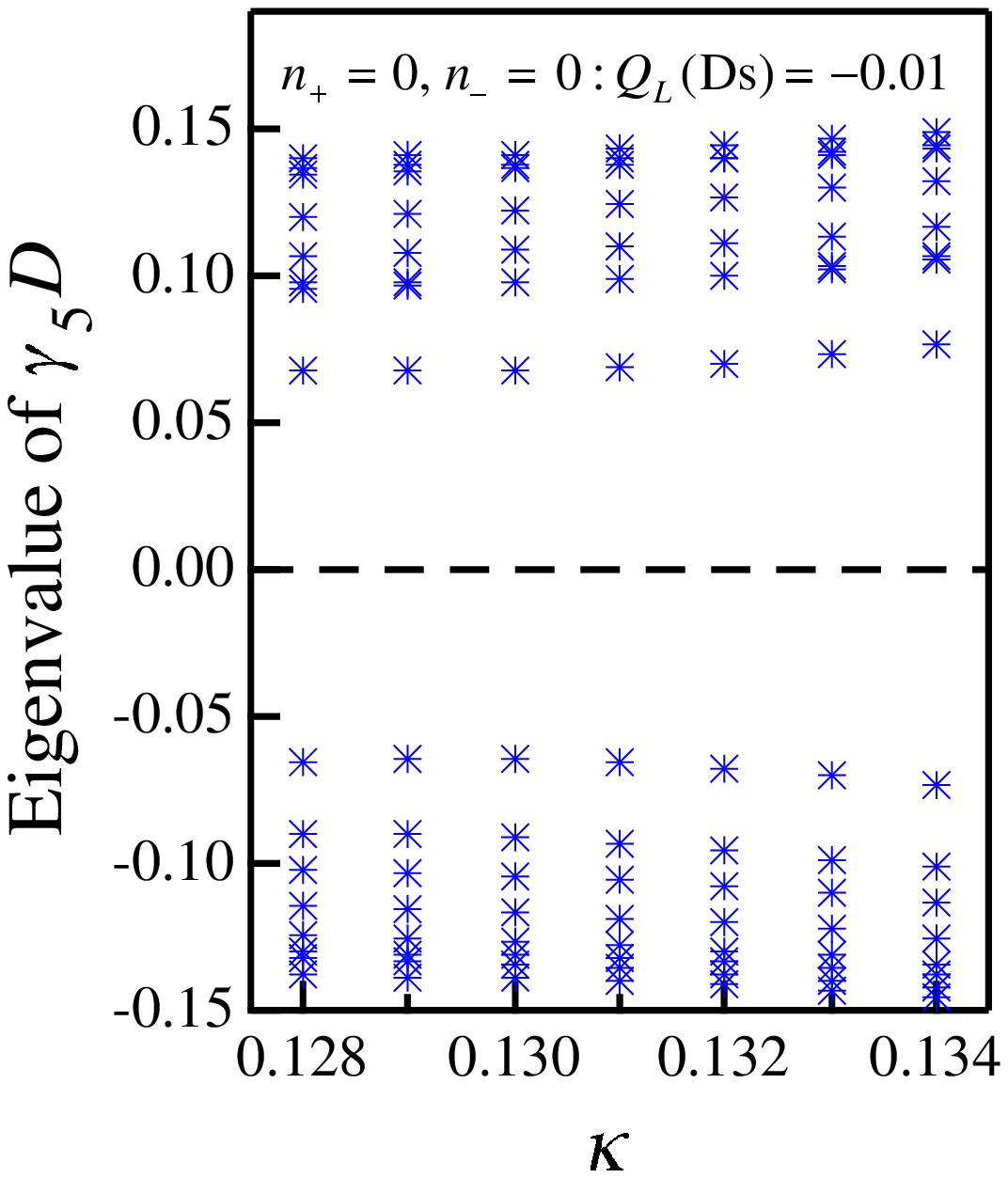}}
\centerline{\fcaption{\label{fig:Zero_Ds7}}}
\end{minipage}
\hspace{\columnsep}
\begin{minipage}{\minitwocolumn}
\centerline{\epsfxsize=7.0cm
\epsfbox{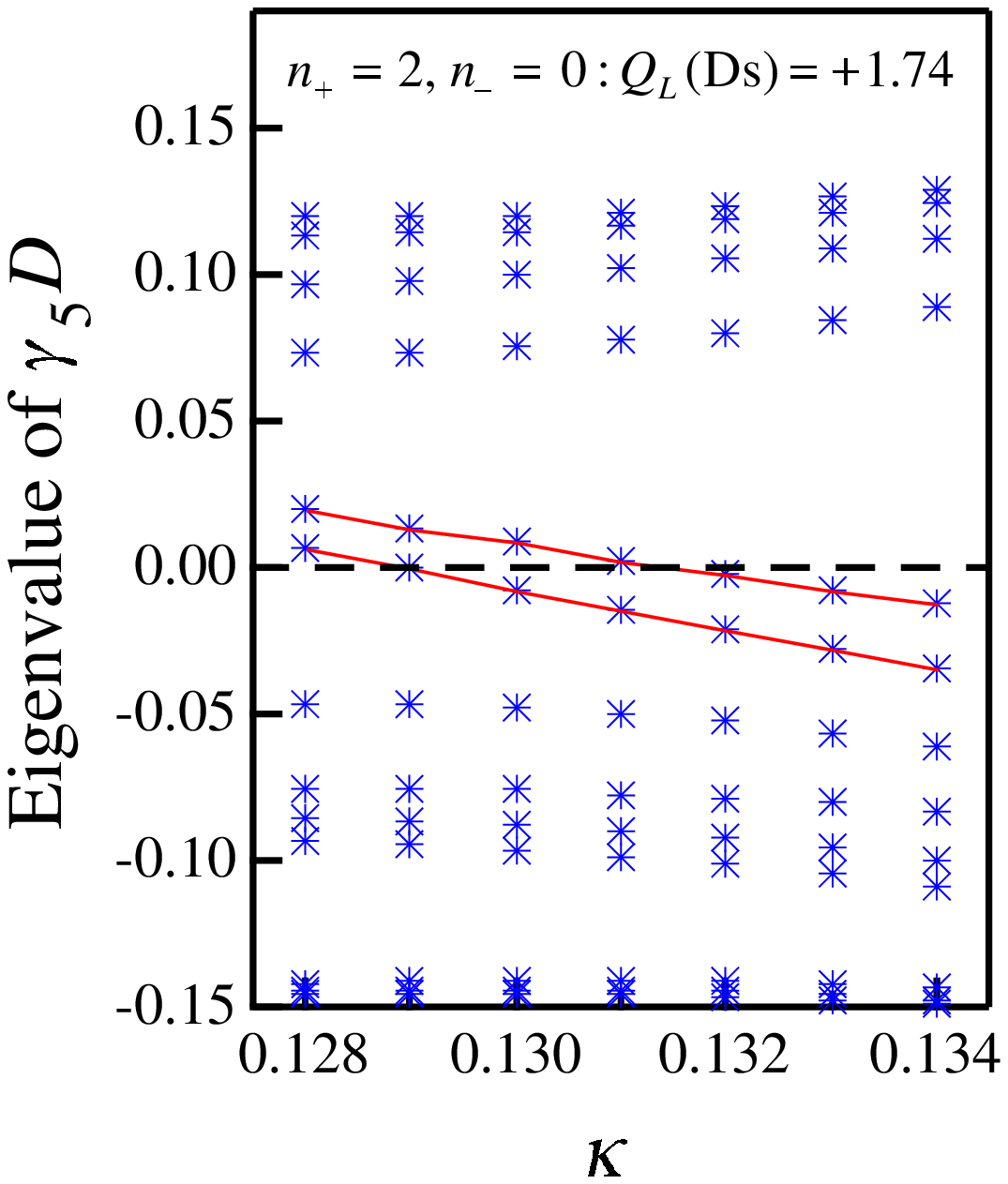}}
\centerline{\fcaption{\label{fig:Zero_Ds8}}}
\end{minipage}
\setcounter{figure}{\value{enumi}}
}
\newpage

\vspace*{1.5cm}
{\setcounter{enumi}{\value{figure}}
\addtocounter{enumi}{1}
\setcounter{figure}{0}
\renewcommand{\thefigure}{\arabic{enumi}(\alph{figure})}

%
%
\noindent
\begin{minipage}{\minitwocolumn}
\centerline{\epsfxsize=7.0cm
\epsfbox{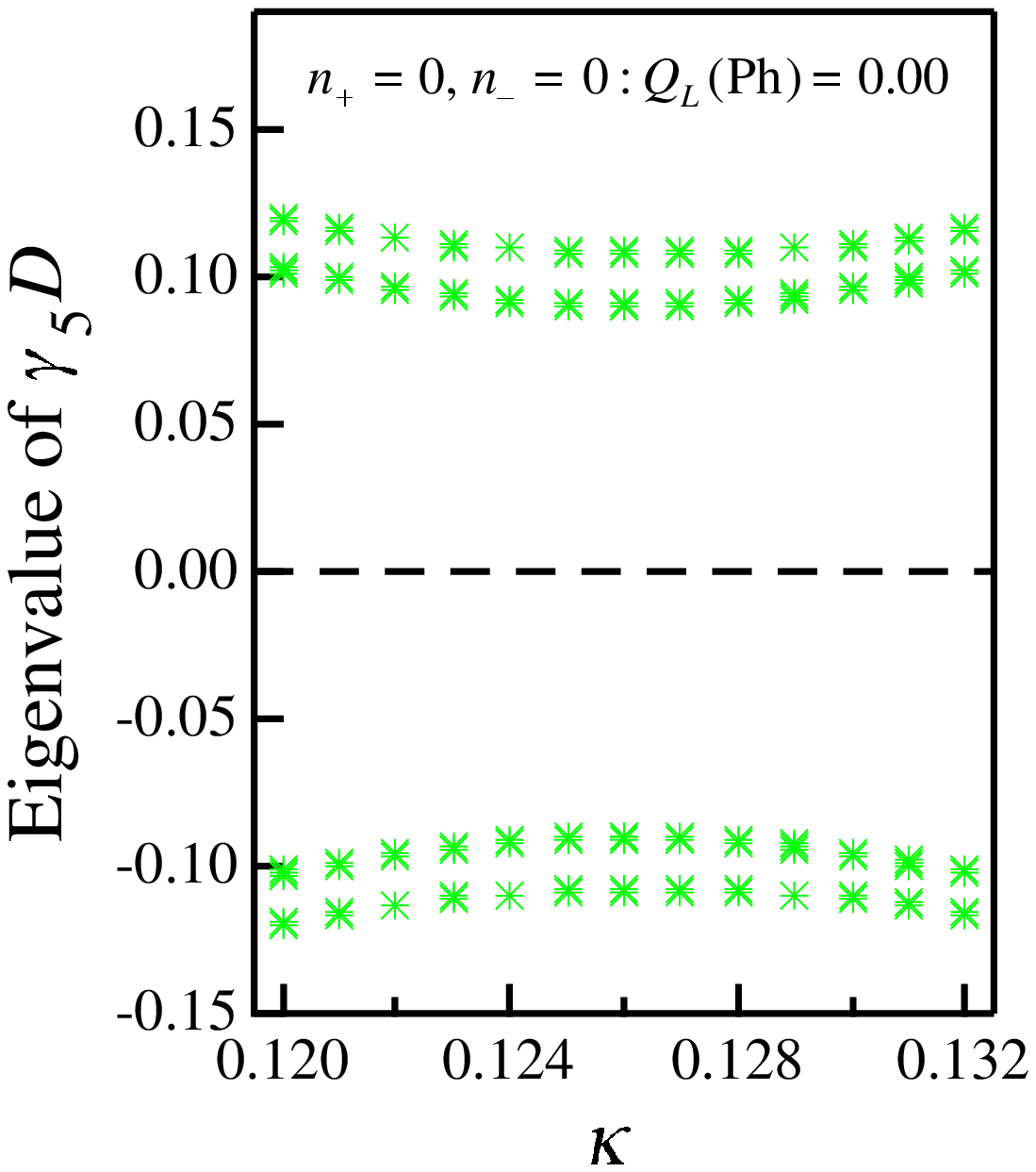}}
\centerline{\fcaption{\label{fig:Zero_Ph1}}}
\end{minipage}
\hspace{\columnsep}
\begin{minipage}{\minitwocolumn}
\centerline{\epsfxsize=7.0cm
\epsfbox{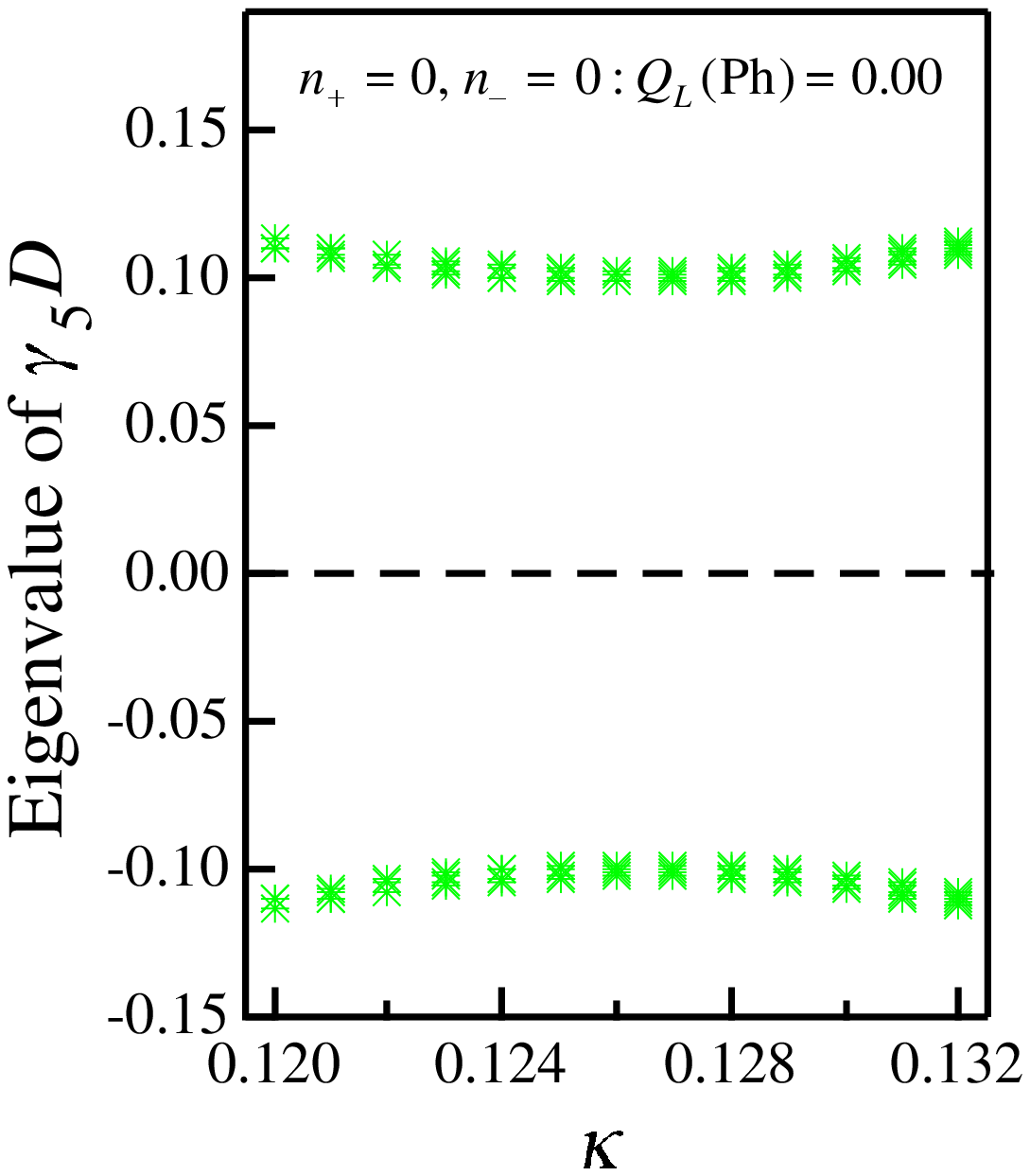}}
\centerline{\fcaption{\label{fig:Zero_Ph2}}}
\end{minipage} 

\vspace*{1.5cm}
\noindent
%
%
\begin{minipage}{\minitwocolumn}
\centerline{\epsfxsize=7.0cm
\epsfbox{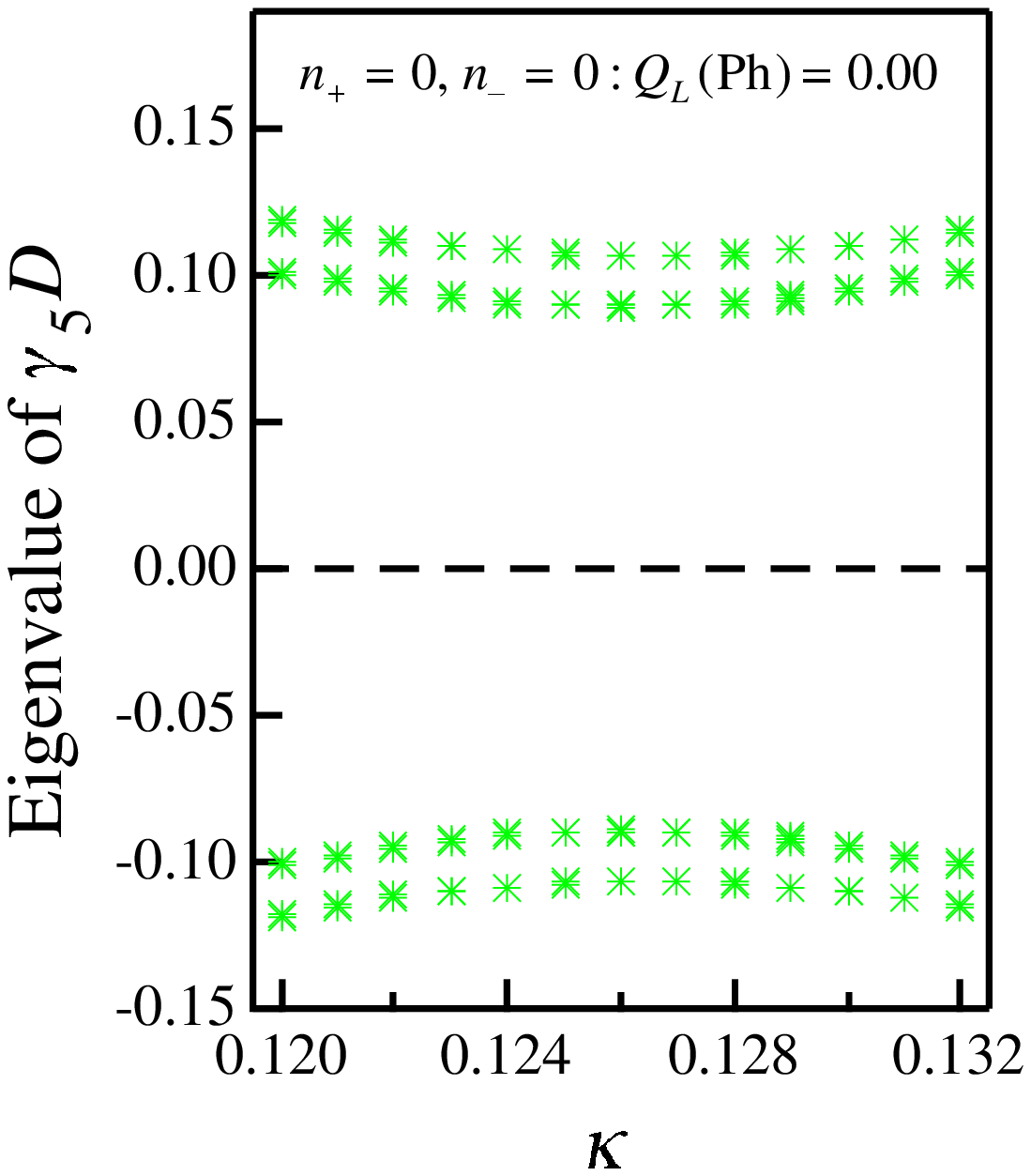}}
\centerline{\fcaption{\label{fig:Zero_Ph3}}}
\end{minipage}
\hspace{\columnsep}
\begin{minipage}{\minitwocolumn}
\centerline{\epsfxsize=7.0cm
\epsfbox{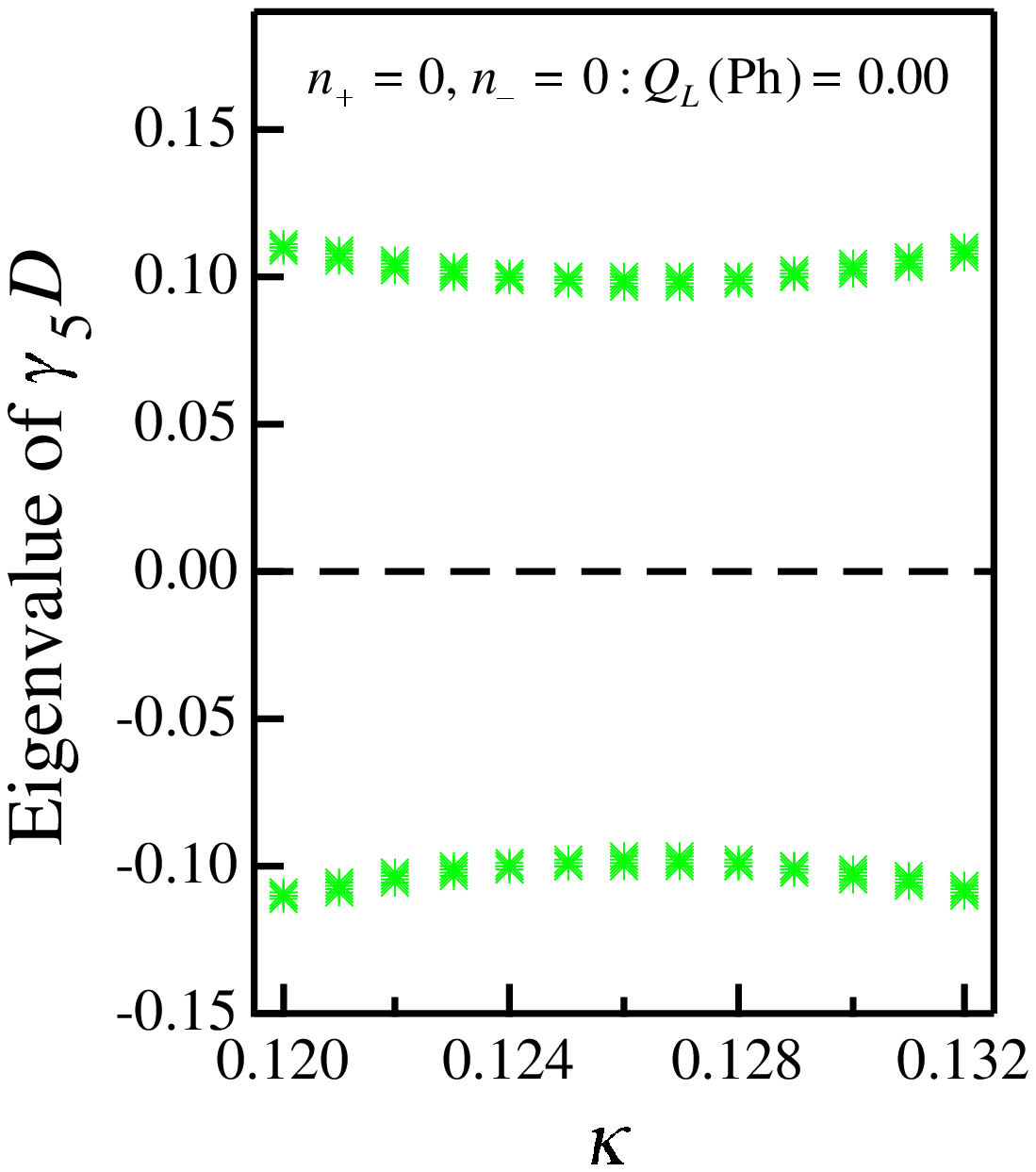}}
\centerline{\fcaption{\label{fig:Zero_Ph4}}}
\end{minipage} 
\newpage

\vspace*{1.5cm}
\noindent
%
%
%
\begin{minipage}{\minitwocolumn}
\centerline{\epsfxsize=7.0cm
\epsfbox{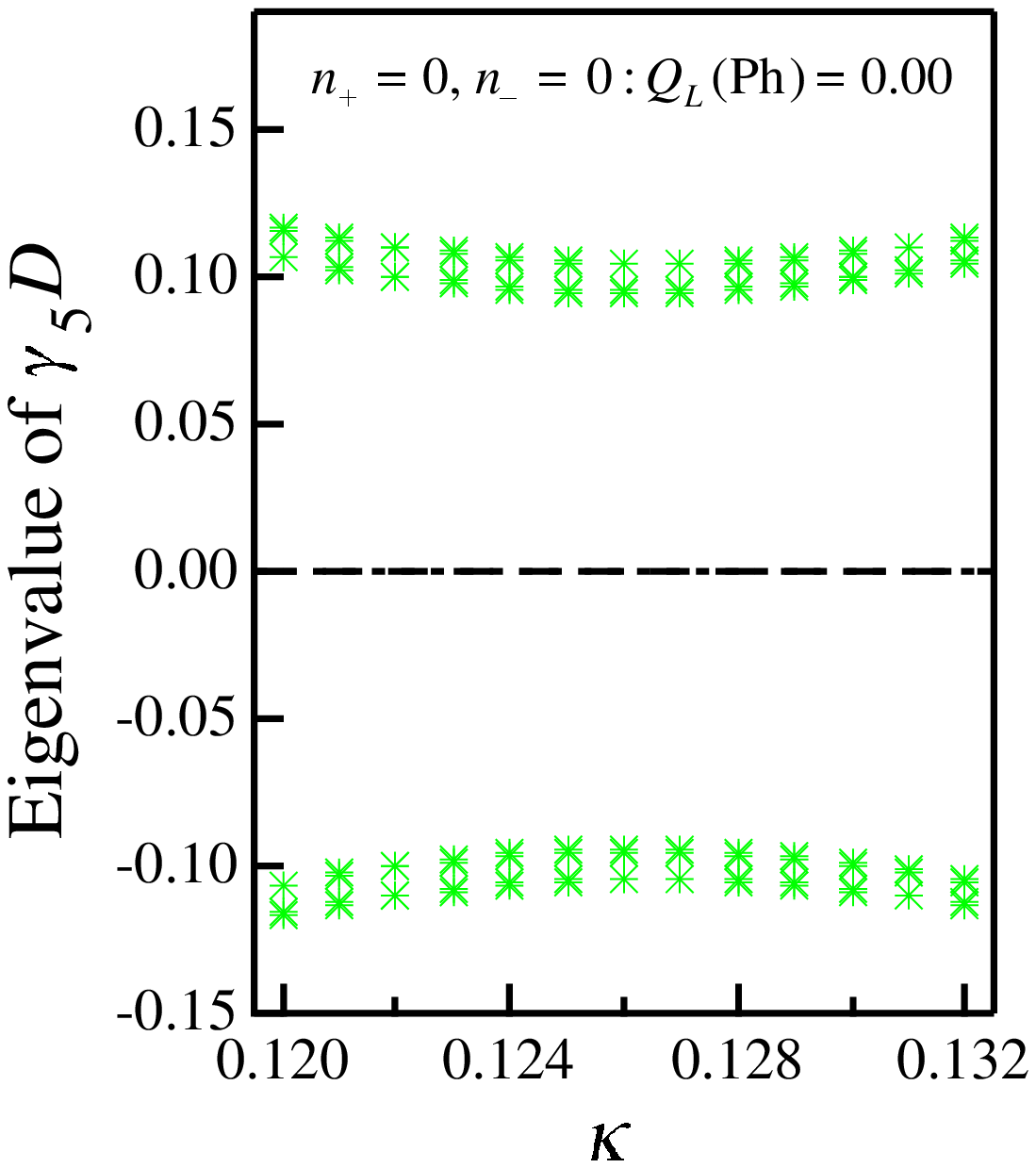}}
\centerline{\fcaption{\label{fig:Zero_Ph5}}}
\end{minipage}
\hspace{\columnsep}
\begin{minipage}{\minitwocolumn}
\centerline{\epsfxsize=7.0cm
\epsfbox{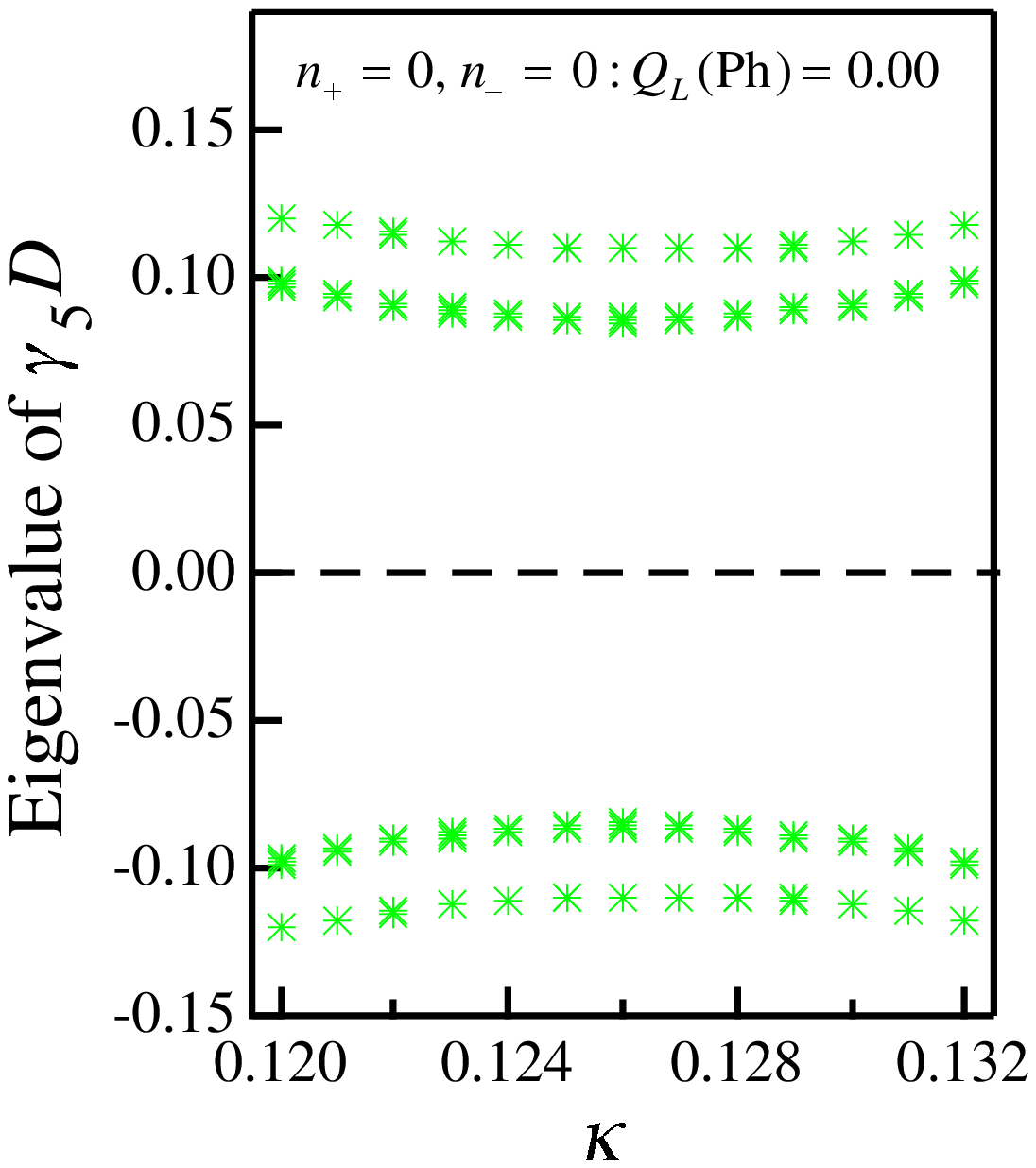}}
\centerline{\fcaption{\label{fig:Zero_Ph6}}}
\end{minipage}

\vspace{1.5cm}
\noindent
%
%
\begin{minipage}{\minitwocolumn}
\centerline{\epsfxsize=7.0cm
\epsfbox{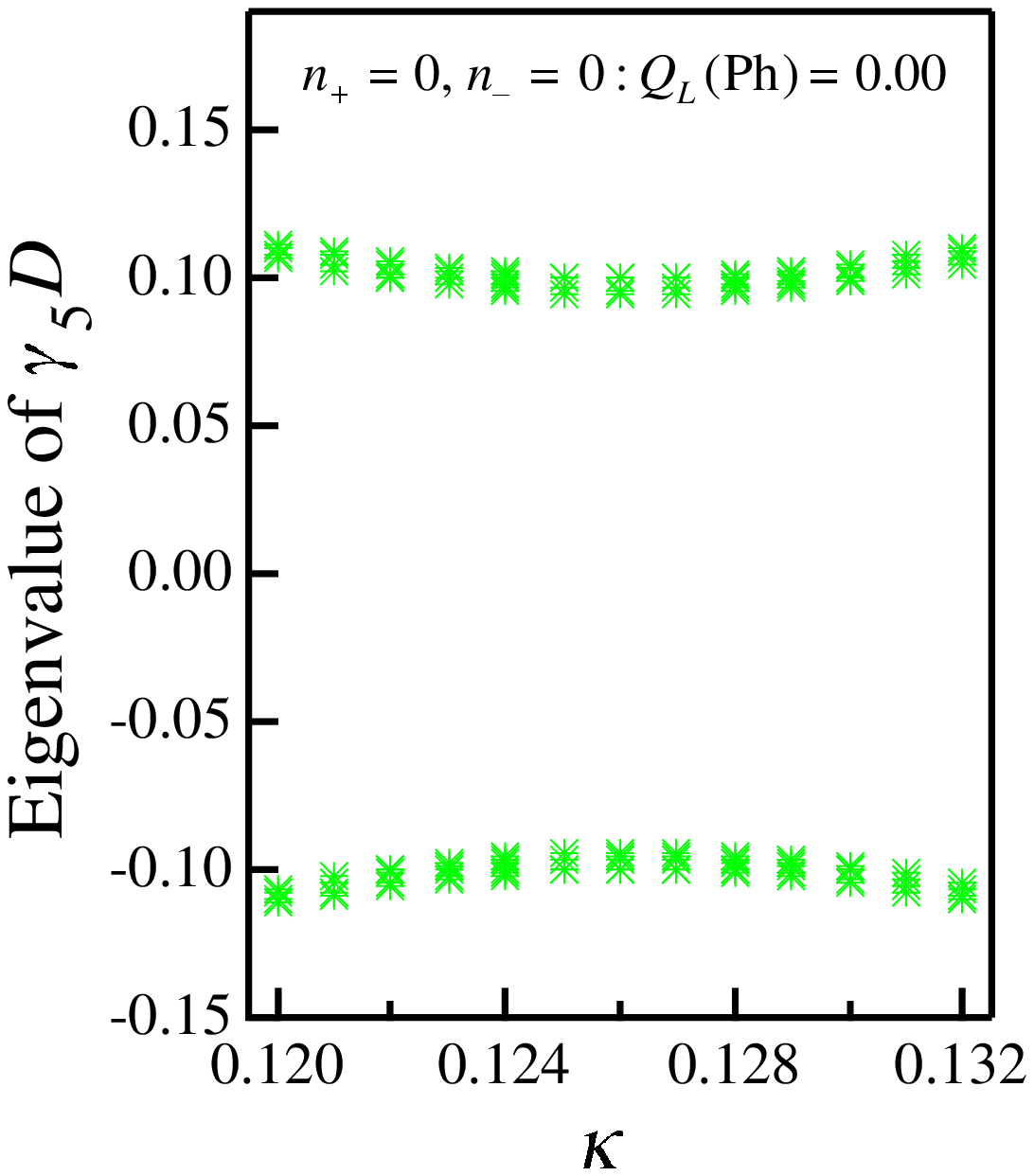}}
\centerline{\fcaption{\label{fig:Zero_Ph7}}}
\end{minipage}
\hspace{\columnsep}
\begin{minipage}{\minitwocolumn}
\centerline{\epsfxsize=7.0cm
\epsfbox{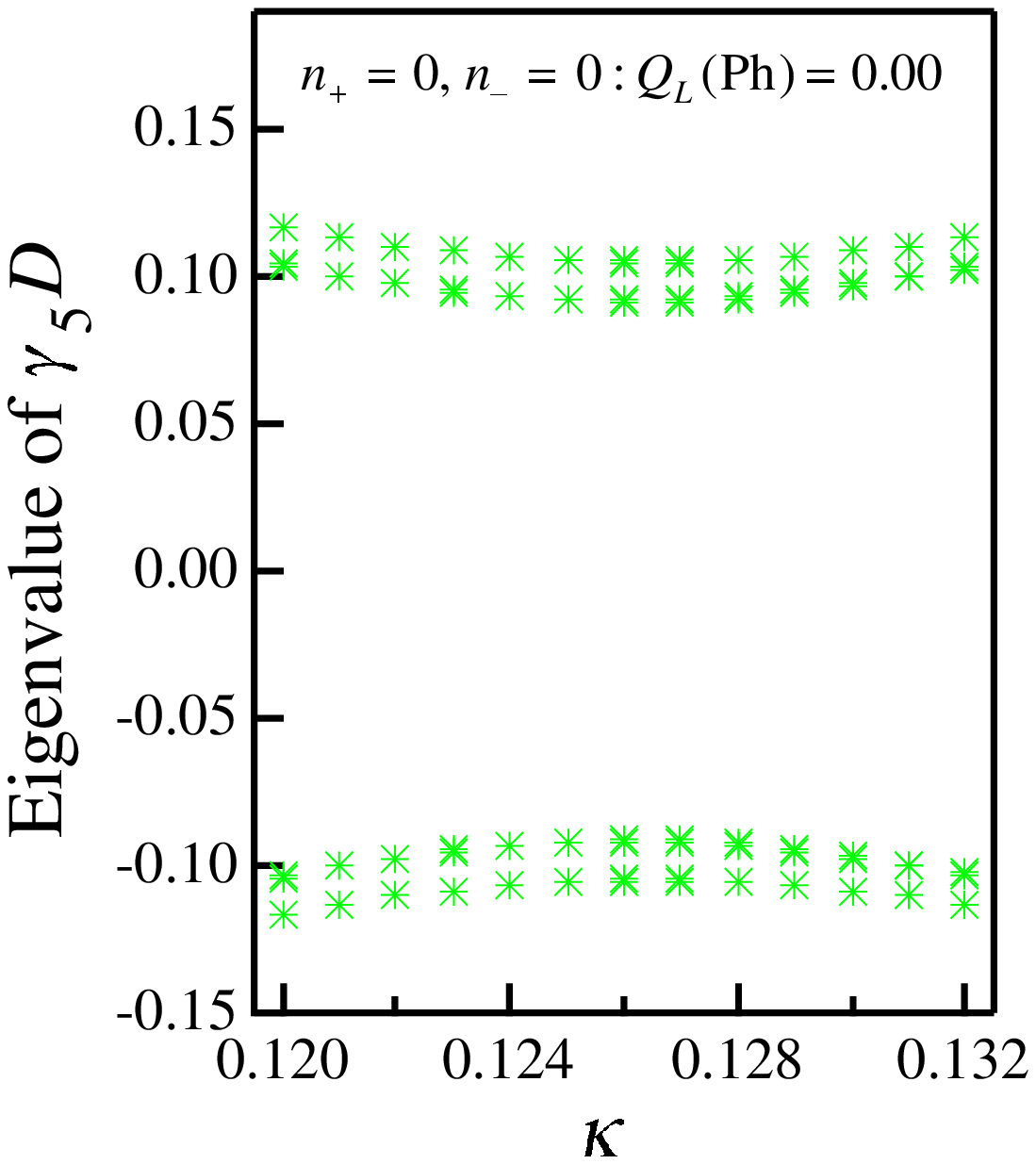}}
\centerline{\fcaption{\label{fig:Zero_Ph8}}}
\end{minipage}
\setcounter{figure}{\value{enumi}}
}
\end{document}